\documentclass[a4paper,11pt]{article}
\usepackage{pos}

\newcommand\CellTopTwo{\rule{0pt}{2.8ex}}
\usepackage{caption} 
\title{VERITAS Dark Matter search in dwarf spheroidal galaxies: an extended analysis}

\ShortTitle{VERITAS DM Search in dSphs: an extended analysis }

\author*[a]{Chiara Giuri}

\affiliation[a]{Deutsches Elektronen Synchrotron DESY, Germany}

 \forColl{VERITAS} 

\emailAdd{chiara.giuri@desy.de}

\abstract{Dark matter (DM) is largely believed to be the dominant component of the matter content of the Universe. Astronomical measurements can be utilized to search for Standard Model (SM) annihilation or decay products of DM, complementing direct and collider-based searches. Among DM particle candidates, Weakly Interacting Massive Particles (WIMPs) are an attractive one. Their decay or annihilation could produce secondary particles including very-high-energy (VHE: $E>100$ GeV) gamma rays, which could be detected by imaging atmospheric Cherenkov telescopes (IACTs). One of the most favourable target classes for DM searches are dwarf spheroidal galaxies (dSphs), dark matter-dominated objects with a negligible predicted gamma-ray emission due to apparent absence of gas and on-going star formation. IACTs, whose point spread functions (PSFs, defined as 68\% containment radius) are typically $0.1^{\circ}$ at 1 TeV, have the necessary angular resolution to detect extended emission from some dSphs. Thus, an extended-source analysis may give an improvement to DM sensitivity, compared to a point-source analysis. In this work, we used observations made since 2007 to 2013 by VERITAS, an array of four imaging atmospheric Cherenkov telescopes sensitive to VHE gamma rays in the 100 GeV - 30 TeV energy range. We performed an unbinned maximum likelihood estimation incorporating the dSph angular profiles of four dSphs and tested its effectiveness against the traditional spectral analysis.}

\FullConference{$37^{th}$ International Cosmic Ray Conference (ICRC2021) \\
		$12^{th} - 23^{rd}$ July, 2021\\
		Online - Berlin, Germany}

\begin{document}
\maketitle

\section{Introduction}
Cosmological observations indicate that visible matter constitutes only around 5\% of the total mass-to-energy content of the Universe. Dark Matter, on the contrary, appears to outnumber visible matter by a factor of six, accounting for around 27\% of the total mass of the Universe~\cite{Riess}. However, we know so little about its nature and formation. An attractive particle candidate for non-baryonic DM is the WIMP.   In the standard scenario, it is believed that WIMPs formed in a pre-BBN era\footnote{Big Bang nucleosynthesis (abbreviated in BBN) predicts the production of the light elements such as D, $
^{3}$He, $^{4}$He and $^{7}$Li, formed during the earliest time of the Universe, i.e. $\sim 200$ s after the Big Bang.} in a thermal plasma, via production and annihilation of WIMP pairs into SM particles and antiparticles. 
Their production ceased when the WIMP annihilation rate became smaller than the expansion rate of the Universe. 
We refer to the final WIMP abundance as the "relic density", whose particles could still annihilate (or decay) into SM products such as gamma rays~\cite{Bertone}. High-energy gamma rays have the advantage of not being deviated by the Galactic magnetic field, so that their arrival directions point directly to the source.

\begin{figure}
\hspace*{-1.4cm}  
\includegraphics[width=1.2\textwidth, height=3.5cm]{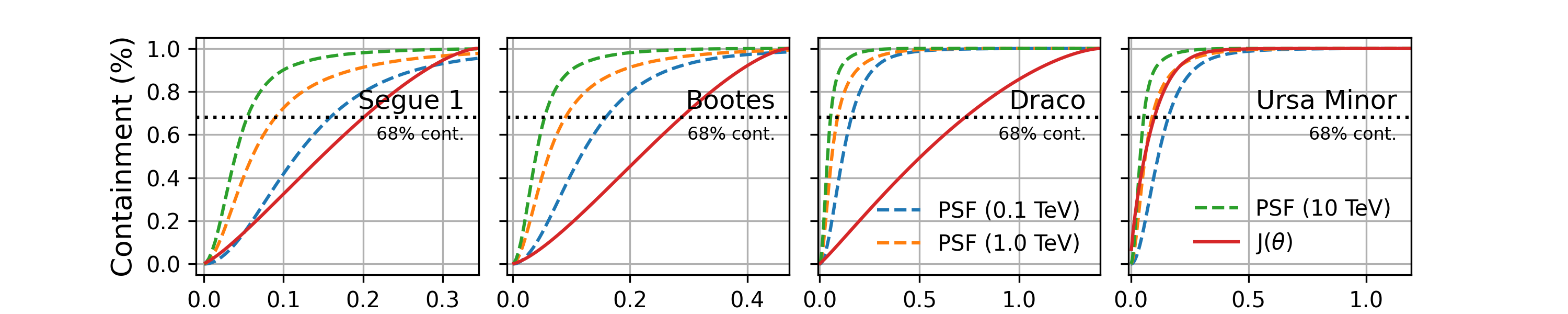}
\caption{The containment fraction for annihilation, represented by the red curve and defined as $J(\theta)/J(\theta_{max})$, as a function of the angular distance $\theta$ [deg] from the center of each dSph. 
The dashed blue, orange and green lines refer, respectively, to VERITAS PSF at 0.1 TeV, 1 TeV and 10 TeV. Containment fraction values taken from~\cite{GS}.}
\label{fig:extension}
\end{figure}

The differential expected $\gamma$-ray flux (in $\gamma \cdot$ m$^{-2} \cdot$ s$^{-1}  \cdot$ GeV$^{-1}$) from DM annihilation can be defined as~\cite{Ahnen}:
\begin{equation}
    \frac{d\Phi}{dE}(\Delta \Omega)=\frac{\langle{\sigma\nu}\rangle}{8\pi M^{2}}\sum_{i}B_{i}\frac{dN_{\gamma,i}}{dE} \times J(\Delta \Omega)\label{eq:flux}
\end{equation}
where the first term is the normalisation factor taking into account the annihilation cross section averaged over the velocity distribution $\langle \sigma\nu \rangle$ and DM particle mass. The second term is the \textit{particle physics factor} including the differential spectrum of each annihilation channel per energy bin $\sum_{i}B_{i}\frac{dN_{\gamma,i}}{dE}$, where $B_{i}$ is the branching ratio into a specific decay. The third term is the \textit{astrophysical $J$ factor} and it describes the DM distribution within the source (see Sec. 2).\\A promising target for the indirect detection of DM are the dwarf spheroidal galaxies, on which this proceeding focuses. They are DM-dominated objects with high mass-to-light ratios~\cite{Strigari}. Located at a comparatively small distance from Earth (about $25-250$ kpc) and at high Galactic latitudes, they do not primarily host star formation regions and are almost gas-free. They have a clean gamma-ray environment, and DM annihilation could easily be correlated with potential gamma-ray emission. Recent studies~\cite{GS} pointed out that some dSphs may be considered as extended sources rather than point-like ones, at least compared to the PSF of some gamma-ray experiments such as VERITAS (see Fig.~\ref{fig:extension}), whose PSF (defined as 68\% containment) is $< 0.1^\circ$ at 1 TeV. So for this reason, including the dSph angular extension as extra information may increase the signal-to-noise ratio and could boost dark matter sensitivity by as much as a factor of two.

In this work four dSphs were studied, i.e. Bo\"otes I, Draco, Segue 1, and Ursa Minor.
We used observations made by the Very Energetic Radiation Imaging Telescope Array System (VERITAS), which is a ground-based observatory that detects gamma-ray photons with energies ranging from 100 GeV to 30 TeV. It is located at the Fred Lawrence Whipple Observatory in southern Arizona, USA (31 40N, 110 57W,  1.3 km a.s.l.). VERITAS consists of four 12-meter telescopes that are spaced 100 meters apart, on average, and it has a typical energy resolution of 15 - 25\%. The flux sensitivity of the standard analysis is such that a source with a flux of order of 1\% of the Crab Nebula flux in the VHE energy range can be detected in approximately $25$ hours of observation~\cite{Park}.

\section{DM distribution in dSphs}
As we already noted, the expected gamma-ray flux from DM annihilation is proportional to the so-called $J$-factor, which gives us information on how DM is distributed within the dSph (see Eq.~\ref{eq:flux} and Fig.~\ref{fig:extension}). In order to infer it, optical stellar-kinematic measurements are used such as the line-of-sight velocity and the position of stars potentially bound to the dSph ~\cite{Mateo}. 
The $J$ factor is defined as: 
\begin{equation}
    \frac{dJ}{d\Omega}=\int_{los}\rho^{2} (l,\Omega)dl
\end{equation}
where $\rho$ is the DM density profile and it is integrated over the line of sight (los) and $\Omega$ is the solid angle\footnote{Defined as $\Delta \Omega=2\pi sin\theta d\theta$.}. Different DM density profiles have been formulated and one of the most corroborated ones is the so-called "generalized" NFW profile ~\cite{Zhao}. It is a five-parameter profile characterizing cold dark matter halos identified in N-body numerical simulations. We adopted this DM profile for this work.

Existing data sets do not provide strong upper constraints on the extent of DM halo, allowing emission to reach to an arbitrarily extended radius. Usually the most conservative choice when determining truncation radius for the DM halo is to choose the outermost member star used to estimate the velocity dispersion profile in the dSph. 
Moreover, due to our imperfect knowledge of the dwarfs and poorly constrained $J$ factors for dSphs, many different realizations of halo $J$-profiles are consistent with the same kinematic data. This leads to systematic uncertainty on the calculation of cross-section upper limits~\cite{Bonnivard}.
In Table~\ref{table:dsphs_data} we show, for the dSphs analysed in this work, some of their most significant properties as well as the integrated $J$ factor values, calculated within a cone of half-angle of 
$\theta_{max}=\arcsin(r_{max}/D)$, where $r_{max}$ is the distance from the center of the dwarf to its outermost member star~\cite{GS}. 

\section{Data analysis and observations}
In this work we analysed VERITAS observations of four dSphs, Bo\"otes I, Draco, Segue 1, and Ursa Minor, observed between 2007 and 2013, resulting in a total quality-selected observation time of $475.65$ hrs.

\begin{table}[t]
\centering
\caption{\textit{On the left}: properties of dSphs, i.e. the distance D from Earth to the center of the dSph, the truncation radius at the outermost observed star $r_{max}$ and the corresponding $\theta_{max}$, the $J$ factor value integrated within a cone of radius $\theta_{max}$ (adopted from~\cite{GS}). \textit{On the right}: VERITAS data analysis results, i.e. total observation time, $N_{\mathrm{on}}$ and $N_{\mathrm{off}}$ counts and the detection significance (in units of standard deviation $\sigma$). Note that the background normalization factor is $\alpha$ = 0.167 for all four dwarfs.}

\small{\begin{tabular}{c|cccc|ccccc}
\hline
\hline
\CellTopTwo{}
Source & Distance & $r_{max}$ &$\theta_{max}$ &$\log_{10}J(\theta_{max})$ 
& Obs.Time & $N_{\mathrm{on}}$ & $N_{\mathrm{off}}$ & $\sigma$\\
& \scriptsize  [kpc] & [pc] &[deg] & \scriptsize{$\log_{10}[\rm{GeV}^2 \rm{cm}^{-5}]$} & [min] & [counts] & [counts] &  \\ 
\hline
\CellTopTwo{}
Bo\"otes I & $66 \pm 2$ & $544^{+252}_{-135}$ & $0.47$ & $18.24^{+0.40}_{-0.37}$ & $950$ & 398 & 2351 & 0.3 \\
\CellTopTwo{}
Draco  & $76\pm 6$ & $1866^{+715}_{-317}$ &$1.30$ & $19.05^{+0.22}_{-0.21}$ & 6813 & 1326 & 8119 & $-0.7$  \\
\CellTopTwo{}
Segue I  & $23 \pm 2$ & $139^{+56}_{-28}$ &$0.35$ & $19.36^{+0.32}_{-0.35}$ & 11042 & 3227 & 19947 & $-1.5$ \\
\CellTopTwo{}
Ursa Minor  & $73 \pm 3$ & $1580^{+626}_{-312}$ &$1.37$ & $18.95^{+0.26}_{-0.18}$ & 9724 & 1328 & 8204 & $-1.5$ \\
\hline
\hline
\end{tabular}}
\label{table:dsphs_data}
\end{table}
We adopted the standard VERITAS reconstruction analysis using the \textit{EventDisplay} software ~\cite{Maier}. We defined a source region (i.e. "ON" region) within $\theta^{2} < 0.008$ deg$^{2}$, where $\theta$ is the angle between the target position and the reconstructed arrival position. In order to reduce the hadronic cosmic-ray background, we applied a
gamma-hadron selection to the data based on \textit{Boosted Decision Trees}~\cite{Krause}. The selection was optimized to give the lowest analysis energy threshold~\cite{Park}. The reflected region model was used to define OFF regions~\cite{Berge}.
Visible starlight can also affect the background estimate, so we removed bright stars by defining circular background exclusion areas centered on stars with minimum apparent magnitudes of $m_{B}<7$ with a size of $0.25^{\circ}$. We excluded also the regions of radius $0.35^{\circ}$ around each dwarf. Table~\ref{table:dsphs_data} shows the data analysis results. For each dSph we show the quality-selected observation time, the  $N_{\mathrm{on}}$ and $N_{\mathrm{off}}$ counts, i.e. the number of observed counts in, respectively, the ON and OFF regions and the detection significance\footnote{Calculated using Eq. 17 in ~\cite{Li}.}. 

\section{Statistical analysis technique}
In order to detect a possible signal from DM annihilation and/or constrain its cross-section, an unbinned maximum likelihood estimation (MLE) can be performed. We adopted a likelihood function including the expected spectral shape from DM annihilation events, which achieves a better sensitivity compared to the conventional likelihood analysis \cite{Rico}.
In this function, in addition to two Poissonian terms, there are two probability density functions to take the likelihood of ON and OFF region events as signal and background into account \cite{Ahnen}:
\begin{equation}
    L=\frac{(g+\alpha b)^{N_{\mathrm{on}}}e^{-(g+\alpha b)}}{N_{\mathrm{on}}!}\frac{b^{N_{\mathrm{off}}e^{-b}}}{N_{\mathrm{off}}!} \prod_{i=1}^{N_{\mathrm{on}}}P_{\mathrm{on}}(E_{i}|M,\langle \sigma_{v} \rangle)\prod_{j=1}^{N_{\mathrm{off}}}P_{\mathrm{off}}(E_{j}),\label{eq:likelihood}
\end{equation}
where $N_{\mathrm{on}}$ ($N_{\mathrm{off}}$) is the number of observed counts in the ON (OFF) region(s), $b$ is the number of expected background counts, and $\alpha$ is the background normalization (i.e., the ratio between number of ON/OFF regions). The parameter $g$ is the total number of expected events from the DM annihilation at a given mass $M$ and averaged annihilation cross-section $\langle \sigma\nu \rangle$, within the ROI: 
\begin{equation}
    \frac{dg}{dE} = \frac{\langle \sigma\nu \rangle T_{obs}}{8\pi M^{2}}\int_{E'}\frac{dN}{dE'} J(E') A(E') D(E|E') dE',
\end{equation}
where $A(E')$ is the effective area as a function of true energy $E'$, $T_{obs}$ is the observation time for a source, $dN/dE'$ is the expected spectrum\footnote{Assuming here a 100\% branching ratio into a certain annihilation final state.} of gamma rays from a DM annihilation event (from~\cite{Cirelli}), $D(E|E')$ represents the probability of an event with true energy $E'$ having a reconstructed energy $E$ (i.e. the energy dispersion matrix), and $J(E')$ is the integrated $J$ factor as a function of true energy $E'$. To compute the integrated $J$ factor, we adopted the NFW DM profile using the median parameters of Table 4 in ~\cite{GS}. We also convolved it with the VERITAS PSF to take into account the finite instrument response. 
In Eq.~\ref{eq:likelihood}, the term $P_{on, i}$ is the likelihood of the $i^{\mathrm{th}}$ ON region event with energy $E_{i}$ to be an event from a distribution composed by the signal from DM annihilation and the background within ROI. The likelihood can be represented by: 
\begin{equation}
    P_{\mathrm{on}, i} (E_{i}|M,\langle \sigma\nu \rangle)=\frac{\alpha b p_{b} (E_{i}) + gp_{g}(E_{i})}{\alpha b + g},
\end{equation}
where $p_{b}$ and $p_{g}$ are the probability density function of background events and expected dark matter signal, respectively, as a function of reconstructed energy. The term $P_{\mathrm{off}}$ is same as $P_{\mathrm{on}}$, but for the OFF region (i.e., $g=0$).

\begin{figure}
\centering{ 
\begin{tabular}{cc}
\includegraphics[width=0.8\textwidth]{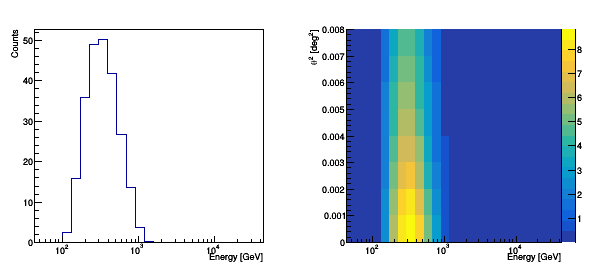}
\end{tabular} 
}
\caption{An example of the expected DM count spectra in the 1D (left) and 2D (right) cases from Segue1 calculated at 1 TeV and assuming $\langle \sigma\nu \rangle=10^{-23}$ cm$^{-3}$s$^{-1}$. In both cases, x-axis corresponds to reconstructed energy and in the 2D histogram the y-axis corresponds to $\theta^2$.}
\label{fig:expected_signal}
\end{figure}

To achieve the main goal of this work, using the angular extension properties of dSphs, we modified the likelihood function in order to include the spatial information in it. Firstly, we computed a two-dimensional (2D) $J$ factor as a function of true energy and solid angle, $J(E', \Omega)$. Also, we built the 2D probability density functions for signal and background; i.e., $p_{g}(E',\theta)$ and $p_{b}(E',\theta)$. Due to the limited background observation, $p_{b}(E',\theta)$ is not continuous and precise; in some very-high-energy bins, there are no background events, so that we cannot assign the background likelihood when events correspond to those bins. To overcome this problem, we fitted the low-energy background distributions with a power law. Then, we extrapolated up to the highest energy we observed in all regions. Note that we confirm that the obtained background model can reproduce the observed background distribution by simulations. The number of expected events from DM annihilation is then given by:
\begin{equation}
    \frac{d^2g}{dEd\Omega}=\frac{\langle \sigma\nu \rangle T_{obs}}{8 \pi M^{2}}\int_{E'} \frac{dN}{dE'}\frac{J(E',\Omega)}{d\Omega}A(E')D(E|E')dE', \label{eq:dg_dOmega}
\end{equation}
whose total number is simply given by integrating Eq.~\ref{eq:dg_dOmega} over the solid angle and the true energy:
\begin{equation}\label{eq:g}
    g = 2\pi \int_{E}\int_{\theta}\frac{d^{2}g}{dEd\Omega} \sin(\theta) dE d\theta.
\end{equation}
Fig.~\ref{fig:expected_signal} shows an example of the expected DM count spectra in the 1D and 2D cases, calculated at 1 TeV and assuming $\langle \sigma\nu \rangle=10^{-23}$ cm$^{-3}$s$^{-1}$. It is evident that counts depend not only on energy but also on $\theta^2$.

For several masses, we minimised the negative logarithm of Eq.~\ref{eq:likelihood} with respect to two free parameters ($\sigma\nu$ and the nuisance parameter $b$) to constrain the DM annihilation cross-section.
The significance of the DM signal was estimated from the likelihood ratio test, TS$=-2(log(L_{0}/L_{1}))$, where $L_{0}$ is null hypothesis (no signal) and $L_{1}$ is the alternative model hypothesis (including signal).

\section{Results and Discussion}
As shown in the final column of Table~\ref{table:dsphs_data}, no DM signal was observed. We studied our sensitivity to a potential signal in the $\tau^+\tau^-$ and $b\bar{b}$ annihilation channels. We performed two simulation studies to address the effectiveness of the 2D analysis, compared to the 1D analysis.

\begin{figure}[h]
\centering 
\centering
\includegraphics[width=0.95\linewidth]{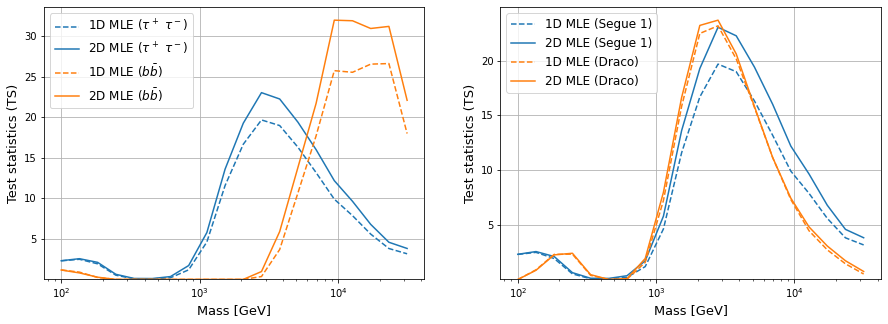}
\caption{Comparing TS as a function of DM mass for 1D and 2D analysis. \textit{On the left}: results for two DM annihilation channels, i.e. $\tau\tau$ and $b\bar{b}$ for Segue1. \textit{On the right}: results for two different dSphs, i.e. Segue1 and Draco. Note that $\langle{\sigma\nu}\rangle$ is set to be different for each channel and each dwarf.}
\label{fig:TS} 
\end{figure}

 Firstly, we checked the effectiveness of the 2D MLE analysis for detecting the DM signal. To test this, we assumed the DM cross-section (high enough to be detected) and produced a simulated on-region distribution, $N_{\mathrm{sim}}(E, \theta) = \alpha N_{\mathrm{off}}(E, \theta) + g(E, \theta)$; for each channel and dwarf, we used different $\langle{\sigma\nu}\rangle$ to make TS values obtained from the MLE be high enough (about 20 to 30)\footnote{For Segue 1, we used $10^{-23.8}$cm$^{3}$s$^{-1}$ ($\tau^+\tau^-)$ and $10^{-22.0}$cm$^{3}$s$^{-1}$($b\bar{b}$). For Draco, we used $10^{-21.6}$cm$^{-3}$s$^{-1}$ ($\tau^+\tau^-)$.}. Then, we synthesized events from the simulated on-region distribution, where the number of the synthesized events is obtained from the Poisson fluctuation of $N_{\mathrm{on}}$. After that, we performed MLE with the synthesized events. From 1000 realizations, we took the average TS value for each mass. Fig.~\ref{fig:TS} shows the comparison between TS values from the 1D and 2D MLE analyses. The left panel of Fig.~\ref{fig:TS} shows that the effectiveness of the 2D method can depend on the DM annihilation channels. Also, the right panel ($\tau^+\tau^-$ for Segue 1 and Draco) shows that the effectiveness of the 2D analysis does not depend on which dwarf is being considered. This implies that the 2D MLE analysis can be more effective in detecting a possible DM signal (Fig.~\ref{fig:TS}) by a factor of 20-30\% (depending on channel and/or source). 


\begin{figure}[!h]
\begin{minipage}[t]{0.45\linewidth}
\centering 
\includegraphics[height=5cm, width =\linewidth]{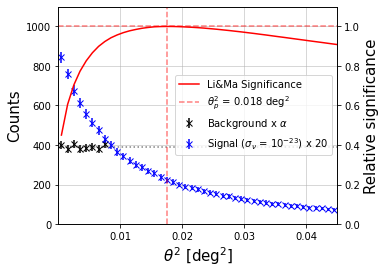}
\caption{Dependence of the Li\&Ma significance on the angular extension for Segue1 at 1 TeV. 
The dashed line is where the Li\&Ma significance peaks ($\theta^2_p$).}\label{fig:LiMa}
\end{minipage}
\hspace{0.5cm}
\begin{minipage}[t]{0.45\linewidth} 
\centering 
\includegraphics[height=5cm,  width =\linewidth]{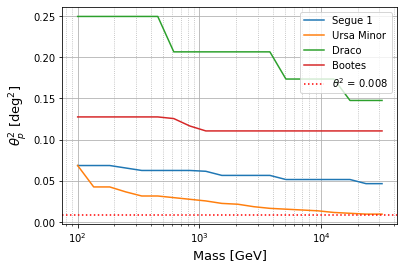}
\caption{Dependence of $\theta^2_p$ on DM mass for the four dSphs considered.} 
\label{fig:optimisation}
\end{minipage}        
\end{figure}

Secondly, we checked how the Li\&Ma significance \cite{Li} varies by extending the size of the source region (Fig.~\ref{fig:LiMa} and Fig.~\ref{fig:optimisation}). For each dwarf and mass, we try to find $\theta^2_{p}$ where the Li\&Ma significance peaks. To do this, we predicted the DM signal count for the $\tau^+\tau^-$ annihilation channel (Eq.~\ref{eq:g}) and calculated the Li\&Ma significance, assuming that $N_{\mathrm{on}} = \alpha N_{\mathrm{off}} + g$ (Fig.~\ref{fig:optimisation}). Above the $\theta^2$ cut ($>$0.008 deg$^2$), we assumed that the background rate is constant. Note that in principle the background rate is independent of $\theta^2$ within the central region of the camera. Fig.~\ref{fig:optimisation} shows $\theta^2_{p}$ as a function of mass. Note that $\theta^2_{p}$ is independent of the DM cross section and exposure time. We can see that the value of $\theta^2_{p}$ tends to decrease as mass increases. This simulation study implies that to obtain better results with the 2D analysis, we need to perform the extended-source analysis with a source region whose size was optimized, depending on the characteristics of the dwarf. 



\section{Conclusions}
In this work, we analysed VERITAS data taken from 2007--2013 on four dSphs for a total quality-selected observation time of 475.65 hrs, using a point-source analysis ($\theta^{2}$ cut $=0.008$ deg$^{2}$). Since dSphs are extended sources, we performed an unbinned maximum likelihood analysis taking into account the angular extension information of the sources. Hence, we derived a likelihood function in which dependencies on both energy and $\theta^2$ were included. We compared the 2D analysis results with those from the traditional 1D analysis (using only energy dependence) to evaluate the effectiveness and usefulness of the angular extension information. 
 With the current $\theta^{2}$ cut, it is unlikely to see any significant improvement in the 2D analysis compared to the 1D method. However, from the simulation studies on the advantage of this 2D MLE analysis over the 1D analysis, if the DM signal exists, it can be inferred that the 2D analysis will be better for detecting and/or constraining its properties. For this reason, we expect that the usage of extended $\theta^{2}$ cuts would lead to improved and more constraining results.


\section{Acknowledgements}
This research is supported by grants from the U.S. Department of Energy Office of Science, the U.S. National Science Foundation and the Smithsonian Institution, by NSERC in Canada, and by the Helmholtz Association in Germany. This research used resources provided by the Open Science Grid, which is supported by the National Science Foundation and the U.S. Department of Energy's Office of Science, and resources of the National Energy Research Scientific Computing Center (NERSC), a U.S. Department of Energy Office of Science User Facility operated under Contract No. DE-AC02-05CH11231. We acknowledge the excellent work of the technical support staff at the Fred Lawrence Whipple Observatory and at the collaborating institutions in the construction and operation of the instrument.

\clearpage \section*{Full Authors List: \Coll\ Collaboration}

\scriptsize
\noindent
C.~B.~Adams$^{1}$,
A.~Archer$^{2}$,
W.~Benbow$^{3}$,
A.~Brill$^{1}$,
J.~H.~Buckley$^{4}$,
M.~Capasso$^{5}$,
J.~L.~Christiansen$^{6}$,
A.~J.~Chromey$^{7}$, 
M.~Errando$^{4}$,
A.~Falcone$^{8}$,
K.~A.~Farrell$^{9}$,
Q.~Feng$^{5}$,
G.~M.~Foote$^{10}$,
L.~Fortson$^{11}$,
A.~Furniss$^{12}$,
A.~Gent$^{13}$,
G.~H.~Gillanders$^{14}$,
C.~Giuri$^{15}$,
O.~Gueta$^{15}$,
D.~Hanna$^{16}$,
O.~Hervet$^{17}$,
J.~Holder$^{10}$,
B.~Hona$^{18}$,
T.~B.~Humensky$^{1}$,
W.~Jin$^{19}$,
P.~Kaaret$^{20}$,
M.~Kertzman$^{2}$,
T.~K.~Kleiner$^{15}$,
S.~Kumar$^{16}$,
M.~J.~Lang$^{14}$,
M.~Lundy$^{16}$,
G.~Maier$^{15}$,
C.~E~McGrath$^{9}$,
P.~Moriarty$^{14}$,
R.~Mukherjee$^{5}$,
D.~Nieto$^{21}$,
M.~Nievas-Rosillo$^{15}$,
S.~O'Brien$^{16}$,
R.~A.~Ong$^{22}$,
A.~N.~Otte$^{13}$,
S.~R. Patel$^{15}$,
S.~Patel$^{20}$,
K.~Pfrang$^{15}$,
M.~Pohl$^{23,15}$,
R.~R.~Prado$^{15}$,
E.~Pueschel$^{15}$,
J.~Quinn$^{9}$,
K.~Ragan$^{16}$,
P.~T.~Reynolds$^{24}$,
D.~Ribeiro$^{1}$,
E.~Roache$^{3}$,
J.~L.~Ryan$^{22}$,
I.~Sadeh$^{15}$,
M.~Santander$^{19}$,
G.~H.~Sembroski$^{25}$,
R.~Shang$^{22}$,
D.~Tak$^{15}$,
V.~V.~Vassiliev$^{22}$,
A.~Weinstein$^{7}$,
D.~A.~Williams$^{17}$,
and 
T.~J.~Williamson$^{10}$\\
\noindent \\
$^1${Physics Department, Columbia University, New York, NY 10027, USA}
$^{2}${Department of Physics and Astronomy, DePauw University, Greencastle, IN 46135-0037, USA}
$^3${Center for Astrophysics $|$ Harvard \& Smithsonian, Cambridge, MA 02138, USA}
$^4${Department of Physics, Washington University, St. Louis, MO 63130, USA}
$^5${Department of Physics and Astronomy, Barnard College, Columbia University, NY 10027, USA}
$^6${Physics Department, California Polytechnic State University, San Luis Obispo, CA 94307, USA} 
$^7${Department of Physics and Astronomy, Iowa State University, Ames, IA 50011, USA}
$^8${Department of Astronomy and Astrophysics, 525 Davey Lab, Pennsylvania State University, University Park, PA 16802, USA}
$^9${School of Physics, University College Dublin, Belfield, Dublin 4, Ireland}
$^{10}${Department of Physics and Astronomy and the Bartol Research Institute, University of Delaware, Newark, DE 19716, USA}
$^{11}${School of Physics and Astronomy, University of Minnesota, Minneapolis, MN 55455, USA}
$^{12}${Department of Physics, California State University - East Bay, Hayward, CA 94542, USA}
$^{13}${School of Physics and Center for Relativistic Astrophysics, Georgia Institute of Technology, 837 State Street NW, Atlanta, GA 30332-0430}
$^{14}${School of Physics, National University of Ireland Galway, University Road, Galway, Ireland}
$^{15}${DESY, Platanenallee 6, 15738 Zeuthen, Germany}
$^{16}${Physics Department, McGill University, Montreal, QC H3A 2T8, Canada}
$^{17}${Santa Cruz Institute for Particle Physics and Department of Physics, University of California, Santa Cruz, CA 95064, USA}
$^{18}${Department of Physics and Astronomy, University of Utah, Salt Lake City, UT 84112, USA}
$^{19}${Department of Physics and Astronomy, University of Alabama, Tuscaloosa, AL 35487, USA}
$^{20}${Department of Physics and Astronomy, University of Iowa, Van Allen Hall, Iowa City, IA 52242, USA}
$^{21}${Institute of Particle and Cosmos Physics, Universidad Complutense de Madrid, 28040 Madrid, Spain}
$^{22}${Department of Physics and Astronomy, University of California, Los Angeles, CA 90095, USA}
$^{23}${Institute of Physics and Astronomy, University of Potsdam, 14476 Potsdam-Golm, Germany}
$^{24}${Department of Physical Sciences, Munster Technological University, Bishopstown, Cork, T12 P928, Ireland}
$^{25}${Department of Physics and Astronomy, Purdue University, West Lafayette, IN 47907, USA}

\end{document}